\documentclass[namedreferences]{SolarPhysics}
\usepackage[optionalrh]{spr-sola-addons} 
\usepackage{graphicx}        
\usepackage{color}           
\usepackage{url}             

\begin{document}
\begin{article}
\begin{opening}

\title{Sensitivity of the g-mode frequencies to pulsation codes and their parameters}

\author{		
A.~\surname{Moya}$^{1}$\sep
S.~\surname{Mathur}$^{2,3}$\sep
R.A.~\surname{Garc\'\i a}$^{3}$ 
         }
\runningauthor{A. Moya {\it et al.}}
\runningtitle{Sensitivity of the g-mode frequencies to pulsation code and its parameters}
\institute{$^{1}$ Departalento de Astrof\'{\i}sica, Laboratorio de Astrof\'isica Estelar y Exoplanetas, LAEX-CAB (INTA-CSIC), PO BOX 78, 28691 Villanueva de la Ca\~nada, Madrid, Spain \url{amoya@cab.inta-csic.es}\\
 	$^{2}$ Indian Institute of Astrophysics, Koramangala, Bangalore 560034, India \url{savita.mathur@cea.fr}\\
   	$^{3}$ Laboratoire AIM, CEA/DSM-CNRS - U. Paris Diderot - IRFU/SAp, 91191 Gif-sur-Yvette Cedex, France \url{rgarcia@cea.fr}\\ 
                      }

\date{Received 5 November 2009; Accepted }

\begin{abstract}

From the recent work of the Evolution and Seismic
  Tools Activity (ESTA, \opencite{2006ESASP1306..363M}; \opencite{2008Ap&SS.316....1L}), whose Task 2
is devoted to compare pulsational frequencies computed using most of
the pulsational codes available in the asteroseismic community, the dependence of the
theoretical frequencies with non-physical choices is now quite well
fixed. To ensure that the accuracy of the computed frequencies is of
the same order of magnitude or better than the observational errors,
some requirements in the equilibrium models and the numerical
resolutions of the pulsational equations must be followed. In
particular, we have verified the numerical accuracy obtained with the
Saclay seismic model, which is used to study the solar g-mode region
(60 to 140$\mu$Hz). We have compared the results coming from the
Aarhus adiabatic pulsation code (ADIPLS), with the frequencies
computed with the Granada Code (GraCo) taking into account several possible choices. We
have concluded that the present equilibrium models and the use of the
Richardson extrapolation ensure an accuracy of the order of $0.01 \mu
Hz$ in the determination of the frequencies, which is quite enough for
our purposes.

\end{abstract}
\end{opening}

\section{Introduction}

The interior of the Sun has been very well studied thanks to
the information provided by pressure-driven modes (p modes). We have
been able to determine some structural variables such as the
sound-speed velocity till the very inner solar core
(e.g. \opencite{2009ApJ...699.1403B}) but with less and less accuracy
towards the deepest layers of the radiative zone. Besides, these
modes are less sensitive to other structural variables like the
density. In the case of the dynamics of the solar interior, due to the
very small number of non-radial p modes penetrating inside the core,
neither the rotation profile (e.g. \opencite{1999MNRAS.308..405C}; \opencite{ThoJCD2003};
\opencite{GarCor2004}; \opencite{2008SoPh..251..119G}) nor the dynamical process (e.g. \opencite{2004A&A...425..229M}; \opencite{2005A&A...440..653M}) are well constrained inside this region.

Gravity (g) modes are eigenmodes propagating inside the Sun while
buoyancy is their restoring force. They can travel inside the
radiative region but when they propagate in the convective zone they
become evanescent reaching the solar surface with tiny amplitudes
(\opencite{2009A&A...494..191B} and references therein). These modes
represent the ``master key'' that would give us a complete access to
the solar core, in particular, to its dynamics
(e.g. \opencite{2008A&A...484..517M}; \opencite{2009arXiv0902.4142M}).

Gravity modes have been searched for a long time, almost since the
beginning of helioseismology (\opencite{HilFro1991};
\opencite{1991AdSpR..11...29P}). But there is currently no undisputed
detection of individual g modes in the Sun
(\opencite{2009arXiv0910.0848A}). However, thanks to the high-quality
observations provided by VIRGO\footnote{Variability of solar IRradiance and Gravity
Oscillations (\opencite{1995SoPh..162..101F})} and GOLF\footnote{Global Oscillations at Low Frequency (\opencite{GabGre1995})} on board SoHO and the ground-based networks BiSON\footnote{Birmingham
Solar Oscillation Network (\opencite{1996SoPh..168....1C})}
 and
GONG\footnote{Global oscillation Network Group (\opencite{HarHil1996})}
, some peaks (\opencite{GabBau2002};
\opencite{2009ApJS..184..288J}) and groups of peaks
(\opencite{STCGar2004}; \opencite{2008AN....329..476G}) have been
considered as reliable g-mode candidates as they have more than
90$\%$ of confidence level and they present several of their expected
properties. Moreover, to increase the probability of detection,
\inlinecite{2007Sci...316.1591G} searched for the global signature of
such modes instead of looking for individual g modes. They found the
signature of the asymptotic dipole g modes with more than 99.99$\%$
confidence level. The detailed study of this asymptotic periodicity
revealed a higher rotation rate in the core than in the rest of the
radiative region and a better agreement with solar models computed
with old surface abundances compared to the new ones
(\opencite{2008SoPh..251..135G}). However, it was not possible to
identify the sequence of individual peaks generating the detected
signal because of the very small signal-to-noise ratio. Thus, to go
further it is necessary to use theoretical g-mode predictions to guide
our search (\opencite{GarciaHofA2010}). For this purpose, we need to know the limits of the
modeled physical processes and quantities as well as the internal
numerical errors of the codes used to compute the predicted
frequencies.

Up to now, the physical processes and quantities included in the solar
models have been improved thanks to the constraints brought by
observations. Many physical inputs have been added or changed such as
the microscopic diffusion, the diffusion in the tachocline or the
chemical composition.  Besides, more studies on these models have been
done since the release of the solar composition of
\inlinecite{2005ASPC..336...25A} leading to a decrease of the
metallicity. The models including the latter composition presented
larger discrepancies in the sound-speed profile
(\opencite{2005ApJ...621L..85B}; \opencite{STCCou2004}) than those
based in the former solar surface abundances
(\opencite{1993A&A...271..587G}). Recently, the decrease of the
metallicity has been reviewed (\opencite{2009arXiv0909.0948A}) and the
differences between the models and the observations are slightly lower
(\opencite{2009ApJ...705L.123S}). The accuracy of the model has already
  been studied in \opencite{2007ApJ...668..594M};
  \opencite{2007A&A...469.1145Z}. They showed that models with
  different physical inputs and fixed surface abundances present
  differences in the frequencies of the g modes that are below 1
  $\mu$Hz in the range [60,140] $\mu$Hz.

In the present paper we study the numerical errors introduced by the
approaches followed by the oscillation codes used to compute the
g-mode frequencies of the Sun. It is a direct application of the study
done in the ESTA group to the solar case and the calculation of the
g-mode frequencies. The main aim of the Task 2 of this group
(\opencite{2008Ap&SS.316..231M}) is to compare different pulsational
codes (a total of nine in this case) when a fixed equilibrium model is
provided, the same for all. Therefore, any differences between the
results provided by these codes are due to non-physical reasons. This
comparison makes it possible to fix the global uncertainties of the
numerical schemes used. The final goal of the group is to provide some
recomendations to ensure that these differences are lower than the
observational accuracies. In this work, we will start by recalling the
methodology followed by the ESTA group in section 2. Then, in Section
3 we will briefly describe the solar-structure model and we will
finish by discussing the results in Section~4.

\section{The ESTA group}

Within the CoRoT Seismology Working Group, the ESTA group (\opencite{2006ESASP1306..363M};\opencite{2008Ap&SS.316....1L}) has been set
up with the aim to extensively test, compare and optimise the
numerical tools used to calculate stellar models and their oscillation
frequencies. Its goals are 1) to be able to produce theoretical
seismic predictions by means of different numerical codes and to
understand the possible differences between them, and 2) to bring
stellar models at the level of accuracy required to interpret the
seismic data from space.

Nine pulsational codes joined the sub-group devoted to the eigenmode-fre\-quen\-cy
comparison. The first step of this sub-group was to test the accuracy
of the results provided by these codes under the different
mathematical schemes and/or algorithms used. To do so, an equilibrium
model was fixed and distributed to the group. Therefore, any eventual
difference could only be due to non-physical differences. To
understand and minimize these differences compared to the
observational uncertainties was the main goal of that work. A $1.5M_\odot$ model was
used, and all the frequency ranges investigated.

The main sources of non-physical differences are the following:

\begin{itemize}

\item Set of eigenfunctions: Use of the Lagrangian or the Eulerian
  perturbation of the pressure ($\delta P$ or $P^\prime$). 

\item Order of the integration scheme: Most of the codes use a
  second-order scheme, but some others have implemented a
  fourth-order scheme.

\item Richardson extrapolation: Some of the codes, using a se\-cond-order
  scheme, have the possibility of using Richardson extrapolation
  (\opencite{1981PASJ...33..713S}) to decrease the truncation error.

\item Integration variable: Two integration variables are used: 1) the
  radius ($r$), or 2) the ratio $r/P$.

\end{itemize}

The main conclusions of the work done by this group are the following (\opencite{2008Ap&SS.316..231M}):
1) if the code uses a second order integration scheme, then the
Richardson extrapolation must be used, 2) the numerical description of
the $\nabla_\mu$ zone must not have any numerical unaccuracy for a
correct description of the g modes and the modes in avoided crossing, 3) at least 2000 mesh points are
neccesary, with an optimal mesh of 4000 points and, 4) the same value
of the gravity constant $G$ must be used to obtain the equilibrium
model and the frequencies.

\section{Modeling the Sun interior: computing p and g modes}

\subsection{Solar model}



We have computed one solar model, which is the Seismic model developed
in Saclay (\opencite{STCCou2001}; \opencite{CouSTC2003}). It is a 1D
model computed with the {\it Code d'Evolution Stellaire Adaptatif et
Modulaire} (CESAM, \opencite{1997A&AS..124..597M}). This model was
tuned to better match helioseismic observations (e.~g. the sound-speed
profile), specially in the radiative region. Another goal of this
model was to predict more accurately the neutrino fluxes.

The most commonly solar model used today is the Standard Solar Model
(SSM), such as model S \-(\opencite{JCDDap1996}). Compared to this
model, the Saclay-Seismic model includes the treatment of the diffusion in
the tachocline prescribed by \inlinecite{1992A&A...265..106S}, while
the opacity and p-p reaction rates have been modified. Concerning the
chemical composition, it uses the abundances table given by
\inlinecite{1993oee..conf...14G}. This model is calibrated in terms of
surface metallicity, luminosity, and radius at the age of 4.6~Gyrs
with an accuracy of $10^{-5}$.

As one of the main aims of this study consists of calculating the
frequencies of g modes, and these modes are mainly confined in
  the inner regions of the Sun, we have computed the model using the
  highest resolution in the core of the Sun (below 0.5~R$_\odot$).

Finally, to calculate the p- and g-mode frequencies we have used two
oscillation codes: The Granada Code (GraCo) and the Aarhus adiabatic
pulsation code (ADIPLS), both being part of the ESTA group.

\subsection{Pulsation codes brief description: Aarhus \& GraCo}

GraCo (\opencite{2004A&A...414.1081M}; \opencite{2008Ap&SS.316..129M})
is a non-radial non-adiabatic linear pulsational code using a second
order integration scheme to solve the set of differential
equations. The code also provides frequencies in the adiabatic
approximation.  This code has been used as a reference code for the
ESTA study. Thus all the possible numerical schemes and algorithms in
the literature have been implemented. Therefore, this code makes it
possible to study the numerical accuracy of the frequencies obtained
with any singular equilibrium model. In adition, this code have been
used to study several g mode pulsatiors (\opencite{2006A&A...450..715R}; \opencite{2006A&A...456..261R}).

ADIPLS (\opencite{2008Ap&SS.316..113C}) is one of the first and most
used adiabatic pulsational codes in the world. It is also based on a
second order integration scheme, and it uses the Eulerian variation of
the pressure as eigenfunction. We have used the relaxation method
where the equations are solved together with a normalization
condition.  The frequencies are found by iterating on the outer
boundary condition. The frequencies have not been estimated
  using the Cowling approximation. Finally, to compute these
  frequencies, we have remeshed the model onto 2400 points and
  extrapolated the parameters bellow 0.05 R$_\odot$.

\section{Results}

Using these two codes, we have obtained the adiabatic frequency
spectrum of the Saclay equilibrium model. A complete comparison
following the work done by the ESTA group has been carried out. For
this purpose, the global characteristics of the numerical resolution
for the ADIPLS code's results have been fixed, that is: the use of the
Richardson extrapolation, $P^\prime$ as eigenfunction, $\ln r$ as
integration variable and $\delta P=0$ as mechanical outer boundary
condition. The frequencies obtained with ADIPLS are our reference
frequencies.

On the other hand, the frequencies with GraCo have been obtained using
all the possible options for the numerical resolution described in
Sec. 2. An additional source of difference has been studied here: the
outer mechanical boundary condition. We have studied the influence of
using $\delta P=0$ or the boundary conditions described in eq. 18.47 of
\inlinecite{1989nos..book.....U}, where the following assumptions
  were used: 1) continuity of the eigenfunctions between the interior
  and a quasi-isothermal atmosphere and, 2) there is no flux coming from outside in the star. This is a physical source of difference, but
its influence is of the order of the rest of the non-physical choices
here studied, and it has never been tested, up to our knowledge, for
solar g modes.

We first show, in Fig. 1, an overview of the complete frequency
spectrum from 60 to 4000~$\mu$Hz. The differences for all the options
except the absence of the Richardson extrapolation are presented in
this comparison. If the Richardson extrapolation is not used, we reach
differences up to 10 $\mu$Hz for the largest frequencies. In this
figure, the main differences are observed with different outer
mechanical boundary conditions, followed by the use of a different
integration variable. Nevertheless, this last choice and the rest of
the non-physical choices show differences in the range $[-0.2,
  0.3]~\mu Hz$, similar to those obtained by the ESTA group using a
2000 mesh points model. Note that, if the same set of non-physical
choices are used with both codes, the differences between the results
related to the way in which the codes solve the equations, are similar
to the change of one of these options.

With this model, we can notice a peak that appears around
  300~$\mu Hz$. In Fig. 2 large separations are depicted as a function
  of the frequency for modes with $\ell=1$ and 2 and frequencies in
  the region [150,500] $\mu$Hz. We see who the peak is coincident
  with a minimum followed of a maximum in the large separtions. This
  use to be a signature of the avoided-crossing fenomenum. This
peak was also present in the ESTA studies, showing a difference of
0.1~$\mu Hz$. This overview of all the spectrum shows that, if the
observational accuracy is 0.1 $\mu Hz$ or lower, we must increase the
number of mesh points in the equilibrium model to ensure that the
theoretical frequencies have a dependancy on non-physical choices
lower than this accuracy.

\begin{figure}[!htb*]  
\centerline{\includegraphics[width=0.80\textwidth,clip=]{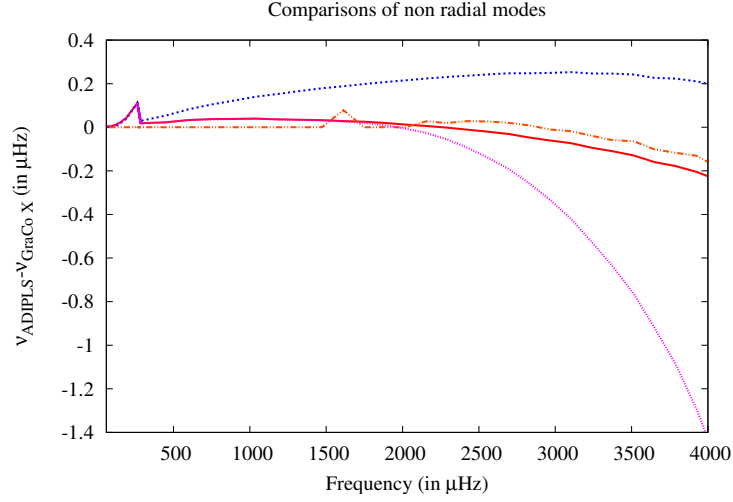} }
             \caption{Overview of the differences found along the
               complete frequency spectrum. The ADIPLS frequencies are
               the reference ones. The differences obtained with
                 all the possibilities explained in the text except
                 the case not using of the Richardson extrapolation are
                 shown (see text for details).}
   \label{Fig1}
   \end{figure}

\begin{figure}[!htb*]  
\centerline{\includegraphics[width=0.80\textwidth,clip=]{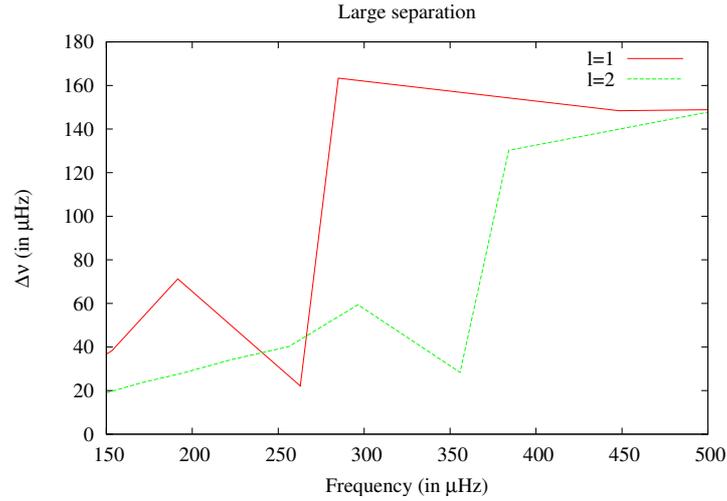} }
             \caption{ Large separations as a function of the frequency
               for modes with $\ell=1$ and 2 in the region around the
               fundamental radial mode.}
   \label{Fig2}
   \end{figure}

Fig. 3 displays the results of the comparisons in the g-mode region
from 60 to 140~$\mu Hz$ for the modes $\ell=1$ and 2, covering
  the radial orders in the ranges n=[4,10] for $\ell=1$ and n=[7,18]
  for $\ell=2$. The results using $\delta P=0$ as mechanical outer
boundary condition are not plotted since it provides, for this
frequency range, the same result as using the
\inlinecite{1989nos..book.....U} boundary condition.

In this figure we see that:

\begin{itemize}

\item The comparison of frequencies for the modes $\ell =1$ and $\ell =2$ provides similar results.

\item When both codes use the same configuration for the numerical
  resolution, the differences found are in the range $[-0.02,0.02]~\mu
  Hz$. These values are much lower than the observational
  accuracy. The reason of these differences must be searched in
    the use or not of the re-mesh. ADIPLS frequencies have been
    obtained using a re-mesh adapted to modes mainly propagating in
    the stellar interior, and GraCo has not this option.

\item When the Richardson extrapolation is not used, the differences obtained are in the range $[0.01,0.08]~\mu Hz$, which can be up to four times bigger than the other comparisons.

\item The rest of the possible choices for the numerical resolution provide differences in the range $[-0.02,0.02]~\mu Hz$. This is similar to the range obtained when the same configuration is used in both codes.

\item The different outer mechanical boundary conditions have no influence in this frequency range.

\end{itemize}

\begin{figure}[htb*]  
 \centerline{\includegraphics[width=0.8\textwidth,clip=]{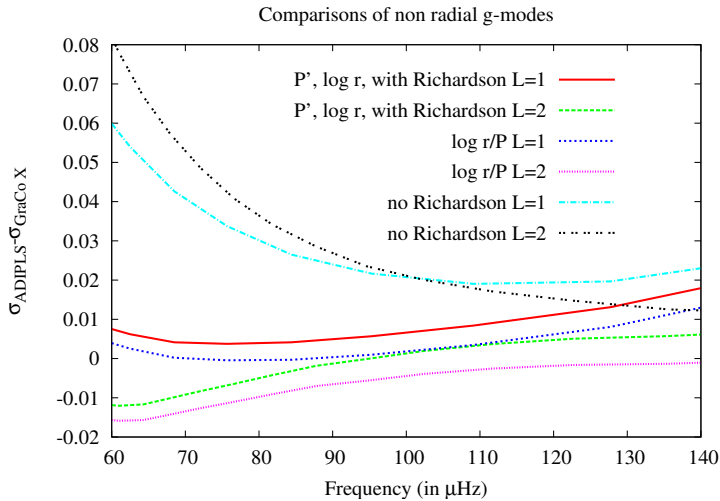} }
             \caption{Differences between the g-mode frequencies from the GraCo code with different options and the reference frequencies from the ADIPLS code as a function of the ADIPLS frequencies in the range $[60,140]~\mu Hz$.}
   \label{Fig3}
   \end{figure}


\section{Conclusions}

The search for g modes has been a long quest as they are the best
probes of the solar core, thus representing a huge potential to better
constrain its structure and dynamics. Up to now, a few candidates have
been detected and recently the global properties of dipole g modes
have been detected with more than 99\% of confidence level. The next step in the search
for individual g modes would consist in being guided by the theoretical predictions of their
frequencies obtained with an oscillation code for a given solar model.

This is the reason why the accuracy of the frequencies calculated with
numerical algorithms is important. In this paper we have taken the
advantage of the previous studies of the ESTA group and we have tested
the accuracy of the equilibrium model used for the search of g modes
in the Sun under changes of non-physical choices in the numerical
integration of the pulsational equations. The Saclay-Seismic model has been
used as an input of the pulsational codes ADIPLS and GraCo. Two
comparisons have been studied: 1) an overview of the differences
obtained along the complete frequency spectrum and, 2) a especial
analysis of the g-mode region $[60,140]~\mu Hz$.

This first comparison has shown that among the
different non-physical choices in several zones of the spectrum, the present equilibrium model
provides differences of the order of 0.1~$\mu Hz$ or larger than that : 
the avoided crossing and the p modes with large frequencies. This
also happens when the same choices are used for both codes. This means
that we need a larger number of mesh points if we want to accurately
fit the observed frequencies in these regions.

On the other hand, the g-mode region presents an accuracy of the order
of $\pm~0.02~\mu Hz$ for any non-physical choice when the Richardson
extrapolation is used which is much better than the uncertainties given by the physical prescriptions used in the models. This study has shown that, if we want to lead a search related to observed g modes in this region and based on theoretical models,
the characteristics of the numerical choices in the Saclay-Seismic model and the use of either ADIPLS or GraCo code, in terms of number of mesh points and numerical accuracy of sentitive quantities, are enough. Thus, such a guided research will be mainly sensitive to uncertainties coming from the physical inputs of the models.




\begin{acks}
The authors want to thank J. Christensen-Dals\-gaard who provided us
the adipack code. AM acknowledges financial support from a {\em Juan de
      la Cierva} contract of the Spanish Ministry of Science and
      Innovation. This work has been partially supported by the CNES/GOLF grant at the Service d'Astrophysique (CEA/Saclay).
\end{acks}

   
\bibliographystyle{spr-mp-sola}

\bibliography{/Users/Savita/Documents/BIBLIO_sav}  

\IfFileExists{\jobname.bbl}{} {\typeout{}
\typeout{****************************************************}
\typeout{****************************************************}
\typeout{** Please run "bibtex \jobname" to obtain} \typeout{**
the bibliography and then re-run LaTeX} \typeout{** twice to fix
the references !}
\typeout{****************************************************}
\typeout{****************************************************}
\typeout{}}

\end{article} 
\end{document}